\documentclass[lettersize,journal]{IEEEtran}
\usepackage{amsmath,amsfonts}
\usepackage{algorithmic}
\usepackage{algorithm}
\usepackage{array}
\usepackage[caption=false,font=normalsize,labelfont=sf,textfont=sf]{subfig}
\usepackage{textcomp}
\usepackage{stfloats}
\usepackage{url}
\usepackage{verbatim}
\usepackage{graphicx}
\usepackage{cite}
\newcommand{\figGA}{Figure~\ref{fig:ga}} 
\newcommand{\figCompFlow}{Figure~\ref{fig:comp}}
\newcommand{\tabSyntheticMetadata}{Table S1} 
\newcommand{\tabSyntheticComplexity}{Table S2} 
\newcommand{\figSyntheticDepthPlot}{Figure~\ref{fig:synDepth}} 
\newcommand{\tabSyntheticProposedComplexity}{Table~\ref{tab:SynComp-lo} and~\ref{tab:SynComp-hi}} 
\newcommand{\tabComplexityGWAS}{Table~\ref{tab:comp-GWAS}} 
\newcommand{\figDepthPlotGWAS}{Figure~\ref{fig:GWASDepth}} 

\newcommand{\tabSummaryGEO}{Table S3} 
\newcommand{\tabComplexityGEO}{Table~\ref{tabcompGeo}} 
\newcommand{\figDepthPlotGEO}{Figure~\ref{fig:GEODepth}} 

\newcommand{\nameOfThisApproach}{\textit{depth}}
\newcommand{\termSizeLimit}{{\tt size\_limit}} 

\begin{document}

\title{How complex is the microarray dataset? \\ A novel data complexity metric for biological high-dimensional microarray data}

\author{Zhendong Sha, Li Zhu, Zijun Jiang, Yuanzhu Chen, and Ting Hu
\IEEEcompsocitemizethanks{\IEEEcompsocthanksitem Zhendong Sha, Li Zhu, Zijun Jiang, Yuanzhu Chen, and Ting Hu are with the School of Computing, Queen's University, Kingston,
Canada, K7L 2N8.\protect\\
E-mail: \{zhendong.sha, 18lz, 19zj6, yuanzhu.chen, ting.hu\}@queensu.ca.}
\thanks{Manuscript received [July] XX, 20XX; revised [Month] XX, 20XX.\\ This work was supported by the Natural Sciences and Engineering Research Council (NSERC) of Canada,
Discovery Grant [RGPIN-2023-03302 to T.H.].\\
(Corresponding author: Ting Hu.)\\
Digital Object Identifier no. \#\#\#\#}
}

\IEEEpubid{0000--0000/00\$00.00~\copyright~2021 IEEE}

\maketitle

\begin{abstract}
Data complexity analysis quantifies the hardness of constructing a predictive model on a given dataset. However, the effectiveness of existing data complexity measures can be challenged by the existence of irrelevant features and feature interactions in biological micro-array data. We propose a novel data complexity measure, {\nameOfThisApproach}, that leverages an evolutionary-inspired feature selection algorithm to quantify the complexity of micro-array data. By examining feature subsets of varying sizes, the approach offers a novel perspective on data complexity analysis. Unlike traditional metrics, {\nameOfThisApproach} is robust to irrelevant features and effectively captures complexity stemming from feature interactions. 
On synthetic micro-array data, {\nameOfThisApproach} outperforms existing methods in robustness to irrelevant features and identifying complexity from feature interactions. 
Applied to case-control genotype and gene-expression micro-array datasets, the results reveal that a single feature of gene-expression data can account for over 90\% of the performance of multi-feature model, confirming the adequacy of the commonly used differentially expressed gene (DEG) feature selection method for the gene-expression data. Our study also demonstrates that constructing predictive models for genotype data is harder than gene expression data. The results in this paper provide evidence for the use of interpretable machine learning algorithms on microarray data.
\end{abstract}


\section{Introduction}
\label{sec:introduction}

\IEEEPARstart{T}{he} increasing availability of micro-array datasets 
has led to the use of machine learning algorithms 
to unveil disease-associated genetic variables~\cite{Libbrecht2015,Nicholls2020}. 
These datasets are structured in the form of {\it X} and {\it y}, 
where {\it X} represents the set of input genetic variables 
that describe the observations, 
and {\it y} represents the corresponding target variable.

While being powerful at exploring disease-associated variables, 
many machine learning algorithms are hard to understand
given their ``black-box" nature of internal architectures.
It is challenging to balance the trade-off between model interpretability and 
predictive capability.
Micro-array data is often difficult to model 
using interpretable approaches, 
owing to the presence of a large number of irrelevant features~\cite{Bauer2014} 
and feature interactions~\cite{Liu2019,ritchie2001multifactor}.
Adopting more powerful ``black-box'' machine learning models 
can compromise the interpretability of the prediction results.
To address this, it is essential to quantify the complexity of a dataset
before constructing a predictive model.
This quantitative understanding helps the design of a prediction algorithm
that maintains sufficient predictive capability
while preserving as much interpretability as possible.

Data complexity metrics 
are used to quantify 
the difficulty of constructing predictive models 
for a given dataset~\cite{Ho2002,Komorniczak2023}. 
These metrics offer 
insights into the complexity of the dataset and 
facilitate the selection of appropriate machine learning algorithms.
Data complexity metrics have been applied in various domains, 
including meta-learning~\cite{vanschoren2018meta}, 
where they provide learning task representation~\cite{Meskhi2021}, 
deep learning, 
where they are used to analyze the effectiveness 
of different layers of neural networks 
in separating classes~\cite{Ho2022}, 
and automation of deep neural networks~\cite{Konuk_2019_ICCV}.
However, the effectiveness of current data complexity metrics on microarray datasets remains to be investigated. 
Different from conventional data, microarray datasets are known 
to possess a large number of irrelevant features 
and exhibit feature interaction.
Feature interaction, or epistasis, 
refers to the phenomenon 
where the effect of one feature on the outcome 
is dependent on the value of another feature~\cite{moore2009}.

\subsection{Existing data complexity measures}
\IEEEpubidadjcol
Data complexity analysis 
is an active research field, 
and existing complexity measures 
can be classified into several types. 
Feature-based measures describe the influence of features 
to separate classes in a classification problem. 
Maximum Fisher's discriminant ratio (\textit{f1}) 
is a measure of overlap between the values of the features 
in different classes~\cite{Ho2002}. 
The directional-vector maximum Fisher's discriminant ratio (\textit{f1v}) 
is a complement to \textit{f1}.
\textit{f1v} quantifies the degree of overlapping between different classes 
by utilizing vector representations 
that have optimized for the separation of classes~\cite{orriols2010documentation}.
The magnitude of the overlapping region (\textit{f2}) 
calculates the overlap of the distribution of feature values. 
The maximum individual feature efficiency (\textit{f3}) 
captures the maximum efficiency of each feature 
in separating the classes~\cite{Ho2002,HO1998}, 
while collective feature efficiency (\textit{f4}) 
quantifies how features work together~\cite{orriols2010documentation}.

Linearity measures quantify 
whether the classes are linearly separable. 
\textit{l1} computes the averaged error distance 
of the miss-classified instances 
to the hyperplane derived from support vector machine (SVM)~\cite{Ho2002,Smith1968}. 
\textit{l2} measures the error rate of an SVM classifier~\cite{Ho2002,Smith1968}, 
while \textit{l3} describes the non-linearity of a linear classifier 
by measuring the classifier's error rate 
on synthesized points of the dataset~\cite{Hoekstra1996}. 
The synthetic points are obtained 
by linearly interpolating instances of each class.

Neighborhood measures characterize 
the presence and density 
of same or different classes in local neighborhoods. 
\textit{n1} creates a minimum spanning tree 
from the instances in the data and computes the percentage of edges connecting data instances from different classes~\cite{Ho2002,Friedman1979}.
\textit{n2} computes the ratio of the sum of the distances 
between each example and its closest neighbor from the same class 
and the sum of the distances between each example and its closest neighbor from different classes~\cite{Ho2002,Friedman1979}. 
\textit{n3} is the error rate of a 1-Nearest-Neighbor classifier~\cite{Ho2002,Friedman1979}, 
while \textit{n4} uses the predictive performance of the NN classifier 
to describe the complexity of the dataset~\cite{Ho2002,Cover1967}. 
\textit{t1} measures the ratio between the number of hyperspheres 
needed to cover the dataset 
and the total number of instances in the dataset~\cite{Ho2002}, 
and \textit{LSC} is calculated as the average cardinality 
between all instances and their nearest instance of different classes~\cite{Leyva2015}.

Network-based complexity measures~\cite{Ho2002} model a dataset as a graph 
and preserve the similarity between instances for modeling the dataset. 
The graph represents instances as nodes and edges are selected based on Gower similarity~\cite{Gower1971}.
\textit{density} measures the number of edges divided by the total possible number of edges in a graph. 
\textit{ClsCoef} measures the density of each node's neighborhood. 
For each node, 
the number of edges between its neighbors is
divided by the maximum possible number of edges. 
The final value of \textit{ClsCoef} is obtained 
by subtracting the sum of all values 
divided by the total number of nodes 
from 1. 
\textit{Hubs} is obtained by subtracting the averaged hub score of nodes in a graph from 1~\cite{Lorena2019}. 
The hub score of node $v_i$ 
is the principal eigenvector of $A^tA$, 
where $A$ is the adjacency matrix of a graph.

Dimensionality measures evaluate data sparsity 
based on the number of samples relative to the data dimensionality. 
\textit{t2} divides the number of instances in the dataset 
by their dimensionality~\cite{basu2006data}. 
\textit{t3} is calculated as the number of components needed for principle component analysis (PCA)
to represent 95\% of data variability divided 
by the number of instances~\cite{LORENA201233}. 
\textit{t4} is calculated as the number of PCA components 
divided by the number of features in the original dataset~\cite{LORENA201233}.

Class imbalance measures 
consider the ratio of the numbers of examples between classes. 
\textit{c1} captures the imbalance in a dataset 
by computing the entropy of class proportions~\cite{LORENA201233}. 
\textit{c2} is a well-known index computed 
for measuring class balance~\cite{Tanwani2010}.

To date, at least three open-source libraries has been published. {\tt DCoL} implements 14 basic complexity metrics in C++ language~\cite{orriols2010documentation}. {\tt ECoL} implements 22 metrics for classification and 12 metrics for regression task in R~\cite{Lorena2018,Lorena2019}. {\tt problexity} implements an equivalent amount of metrics in Python language~\cite{Komorniczak2023}.

We identify two major limitations of existing complexity measures. 
First, metrics that rely on machine learning models 
are not robust to irrelevant features. 
This may lead to inaccurate data complexity estimates. 
Second, there is currently no metric available 
to directly characterize feature interaction.
Although several metrics exist for measuring linearity, 
but none directly addresses feature interaction. 
These limitations make it difficult 
to accurately describe the complexity of high-dimensional micro-array datasets 
that contain feature interactions~\cite{Liu2019}.

\begin{figure*}[h]
    \centering
    \includegraphics[width=.8\linewidth]{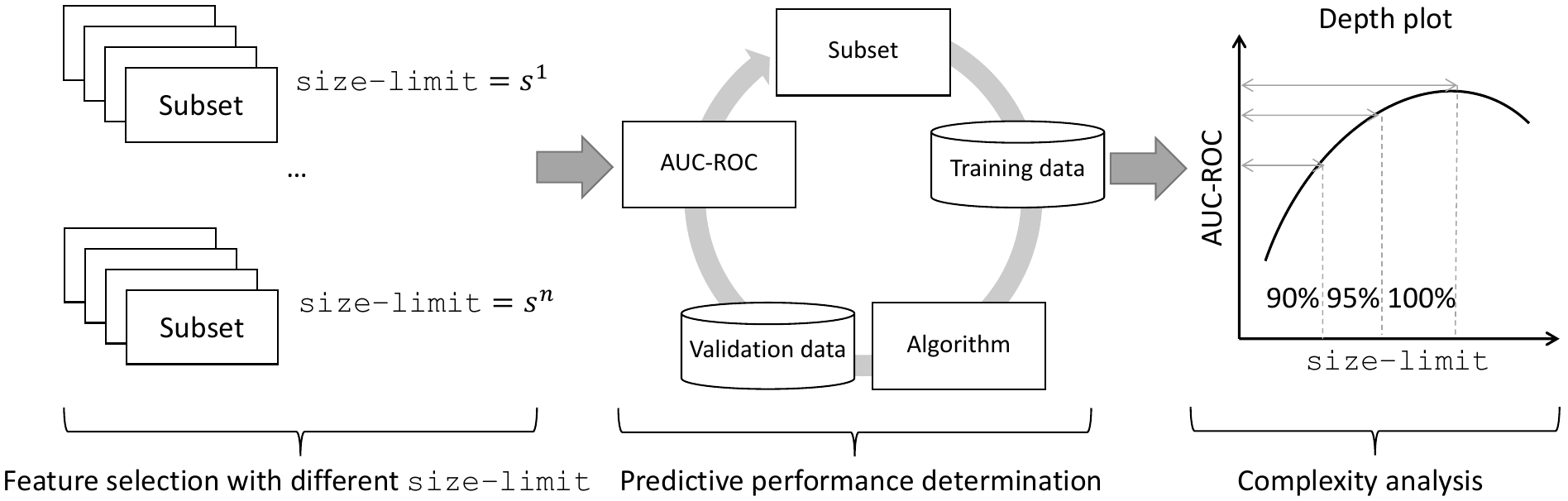}
    \caption{Flowchart of the prediction problem complexity quantification procedure. Several cohorts of feature selection runs (based on {\figGA}) are executed using varying ${\termSizeLimit}=\{s^1,...,s^n\}$ values. The resulting feature subset's predictive performance is then evaluated using the validation data. Since the data complexity is generated from the depth plot, so we refer to this method as {\nameOfThisApproach}.}
    \label{fig:comp}
\end{figure*}

\subsection{Existing feature selection methods}
We intend to leverage feature selection method 
to enable data complexity metrics 
to account for irrelevant features and feature interactions.
Feature selection can extract a subset of relevant features 
from data to construct predictive models. 
Existing feature selection approaches 
can be categorized into three types and 
have been utilized to identify feature interaction. 

The filter approach involves 
using metrics to rank features 
based on their correlation with the target label. 
The top-ranked features are then selected for model construction. 
In micro-array data analysis, 
the most commonly used filter approach 
is differently expressed genes (DEG)~\cite{Anders2010}. 
This approach selects features 
with the highest uni-variable correlation 
with the target label.

The wrapper approach performs feature selection 
by using a machine learning model 
to evaluate the predictive performance of feature subsets. 
Evolutionary algorithms, such as genetic algorithm (GA), 
are popular search strategies for wrapper approaches~\cite{siedlecki1993note,al2017examining,swerhun2020summary,sayed2019nested,garcia2020unsupervised}. 
Various machine learning algorithms, 
including linear regression~\cite{leardi1992genetic}, logistic regression~\cite{Sha2021}, Naive Bayes~\cite{da2011improving,canuto2012genetic,sousa2013email}, support vector machines (SVM)~\cite{sayed2019nested,canuto2012genetic,seo2014feature}, and artificial neural networks (ANN)~\cite{winkler2011identification,souza2011co,oreski2014genetic}, 
are used to evaluate the goodness of a feature subset. 
Different evaluation algorithms 
can lead feature search algorithms 
to discover features with different types of association. 

The embedded approach 
performs feature selection while constructing a predictive model. 
Random forest~\cite{breiman2001random} is a popular learning method with embedded feature selection~\cite{dorani2018ensemble,PHMA13,MoAW10}. 
During the construction of the forest, 
irrelevant features are discarded. 
Regularization techniques, such as Lasso or Ridge regression~\cite{Mak2017}, 
can automatically select the most important features 
by incorporating the loss function 
to penalize the inclusion of less important features.

Feature selection can also be utilized to identify epistasis. 
For instance, multifactor dimensionality reduction (MDR)~\cite{ritchie2001multifactor} 
has been used to search for combinations of genetic variables 
associated with epistasis. 
Random forest has also been applied to genetic data analysis 
to identify feature interaction~\cite{dorani2018ensemble}. 
Evolutionary computation approaches, such as ant colony optimization~\cite{wang2010antepiseeker} and differential evolution~\cite{storn1997differential}, 
have been utilized as search strategies 
to identify epistasis in high-dimensional data~\cite{cao2019hissi,guan2021differential}.

\subsection{The proposed approach}
We introduce a novel method named {\nameOfThisApproach} 
to evaluate data complexity within biological micro-array datasets. 
This approach effectively tackles the challenges 
associated with irrelevant features and feature interaction 
by investigating the predictive performance of feature subsets 
with varying size limits (\figCompFlow). 
We employ feature selection algorithms (Section~\ref{sec:GA})
to detect irrelevant features. 
Additionally, we address data complexity
arising from epistasis 
by performing a non-linear feature selection 
based on decision tree fitness evaluation (Section~\ref{sec:fitness}). 
In the subsequent sections, 
we present the proposed data complexity metrics 
in Section~\ref{sec:complexity}. 
To demonstrate the effectiveness of our approach, 
we first apply it to synthetic data (Section~\ref{sec:syn}), 
followed by an analysis of real high-dimensional micro-array data (Section~\ref{sec:gwas} and~\ref{sec:geo}). 
Finally, we summarize our experimental results 
and discuss the implications of {\nameOfThisApproach} in Section~\ref{sec:conclusion}.

\section{Materials and Methods}

\label{sec:method}
The application of existing data complexity measures 
to biological micro-array datasets 
is very challenging 
due to the high dimensionality and the presence of epistasis. 
To address these challenges, 
we propose a novel data complexity measure {\nameOfThisApproach}
based on a feature selection method. 
This approach is designed with 
an improved robustness to irrelevant features and feature interactions, 
as outlined in {\figCompFlow} and Section~\ref{sec:complexity}.

\subsection{Feature selection based on genetic algorithm}
\label{sec:GA}

Feature selection, which selects a subset of relevant features, 
seeks to optimize the performance of a predictive model. 
Identifying the optimal feature subset 
is a combinatorial optimization problem 
that can be solved using a genetic algorithm (GA), 
a heuristic optimization strategy inspired by biological evolution~\cite{Holland1992}. 
GA has been demonstrated to be effective 
in finding near-optimal feature subsets for high-dimensional data.

In GA, a feature subset is represented as a binary vector of length $d$, 
where $d$ is the total number of features in the data,
and the value of each element of the vector indicates 
whether a feature is selected (1) or not (0). 
GA performs feature selection 
by generating a population of random feature subsets, 
each of which is evaluated to determine its overall relevancy 
for predictive model construction. 
Feature subsets with better performance 
have an evolutionary advantage and 
are more likely to be used for subset reproduction. 
Increasing the number of subsets in GA 
enhances its search capability 
but requires greater computational resources. 
The feature subset reproduction process includes:
\begin{enumerate}
    \item Fitness evaluation: Each feature subset is evaluated for fitness using the mean testing area under the receiver operating characteristic curve (AUC-ROC) of five-fold cross-validation based on a machine learning algorithm (explained in Section~\ref{sec:fitness}).
    \item Selection: The subsets with higher fitness values are selected from the population as parents for the next generation using tournament selection~\cite{miller1995genetic}.
    \item Crossover: Each pair of parental subsets exchange their selected features randomly.
    \item Mutation: Some feature subsets may undergo bit-flip mutation, which either adds or removes a feature.
\end{enumerate}

The GA process (see {\figGA}) continues 
until a specified number of generations are reached, 
and the fittest feature subset is outputted. 
The performance of GA 
depends on several parameters 
such as population size, tournament selection size, and the probabilities of crossover and mutation. 
Thus, parameter tuning is necessary to achieve better result.

\begin{figure}[t]
    \centering
    \includegraphics[width=\linewidth]{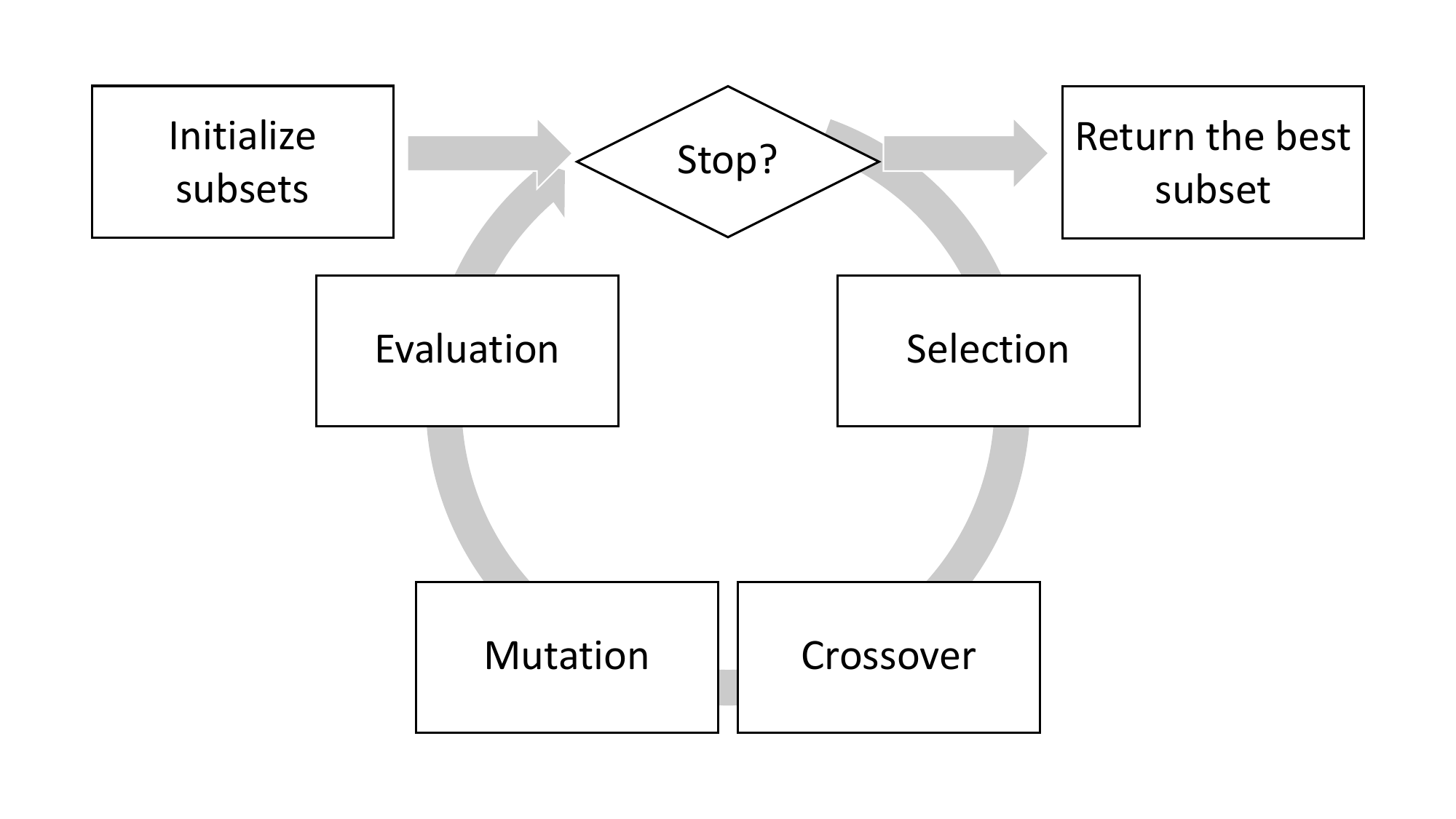}
    \caption{Flowchart of the GA algorithm. }
    \label{fig:ga}
\end{figure}

\subsection{Fitness evaluations}
\label{sec:fitness}

GA evaluates a feature subset based on the prediction performance of a machine learning model that uses the feature subset.
The selection of such a machine learning algorithm will result in selecting features that contributing to the prediction in different ways. 
For instance, 
the use of an additive model 
holds the assumption 
that the influence of each feature on the target variable 
is independent of others~\cite{fisher_1919,Zhou_2022}. 
This section introduces two fitness evaluations 
based on different machine learning algorithms. 

Logistic regression~\cite{Cox1958}, 
a linear model often deployed in polygenic risk predictive models~\cite{choi2020tutorial}, 
aims at discerning the polygenic association 
between features and the target variable. 
Thus, the fitness evaluation 
underpinned by logistic regression 
prioritizes features demonstrating a high uni-variable association with the target.

In contrast, a decision tree 
is a non-parametric model 
capable of capturing non-linear interactions between features~\cite{breiman2017classification}. 
The objective of the decision tree-based fitness evaluation 
is to exploit the decision tree's capacity 
to capture the interactions among features present in a dataset. 
Unlike logistic regression, 
fitness evaluation using the decision tree algorithm 
selects features demonstrating 
both uni-variable association and non-linear feature interaction.

Based on these two fitness evaluations, 
{\nameOfThisApproach} can detect data complexity 
originating from epistasis. 
In the following sections, 
the evaluations based on logistic regression and decision tree algorithms 
will be referred to as {\it linear} and {\it non-linear} evaluations, respectively.

\subsection{Depth plot and prediction problem complexity quantification}
\label{sec:depthPlot}
\label{sec:complexity}

The potential of the GA-based feature selection algorithm 
to discern relevant features 
depends on the number of features $d$ included in each subset. 
We utilize a parameter, named {\termSizeLimit}, 
to control the maximum number of features in each subset,
which can be expressed as $d \leq {\termSizeLimit}$.
By modulating {\termSizeLimit}, 
the predictive performance of the resulting feature subset 
can be used to describe the complexity of the data. 
The correlation between {\termSizeLimit} and predictive performance is illustrated in a depth plot, 
wherein the horizontal axis indicates the {\termSizeLimit} values 
and the vertical axis represents the predictive performances of the feature subsets.


The depth plot is generated from multiple cohorts of feature selection runs with varying {\termSizeLimit} values ({\figCompFlow}).
By gradually increasing the {\termSizeLimit} while retaining the population size constant, 
the change in the predictive performance of feature subsets 
with different {\termSizeLimit} values can be examined. 
For each {\termSizeLimit} value, 
the GA algorithm is iterated multiple times, 
and the predictive performances of the resulting feature subsets 
are summarized within the ``depth plot''. 
The average predictive performance of subsets across various {\termSizeLimit} values
is delineated by a curve.


The slope of the curve in the depth plot 
is used to characterize the complexity of the prediction problem 
for a given dataset. 
Owing to this reason, we name the proposed approach {\nameOfThisApproach}. 
We first identify the {\termSizeLimit} value 
yielding the best averaged predictive performance (100\%). 
Then, the minimum {\termSizeLimit} values capable of achieving 90\%, 95\%, 99\%, and 99.5\% of the peak performance are determined. 
Additionally, we identify the elbow point~\cite{Tolsa2000} in the curve 
where further increasing the {\termSizeLimit} does not significantly improve the predictive performance.

To address overfitting, 
we evaluate the predictive performance of a feature subset 
using both the observations used for feature selection runs and 
unseen observations. 
The dataset is partitioned into three sections: 
training (60\%), testing (20\%), and validation (20\%).
The training and testing folds are employed for feature selection, 
the testing fold is used to assess the predictive performance of the resulting feature subsets on known data,
and the validation fold is used to access the predictive performance on unseen data. 
To evaluate a feature subset,
a predictive algorithm is initially constructed 
based on the training fold. 
Following this, its predictive performance 
is evaluated across all three folds 
to characterize its predictive performance. 
The overall complexity quantification procedure is visualized in {\figCompFlow}.

\subsection{Dataset preparation}
In this study, we aim to assess the efficacy of our proposed methodology across three data types: synthetic data, colorectal cancer genotype data, and gene expression data.

\subsubsection{Synthetic datasets}

To evaluate the effectiveness of our method, 
we utilize six pre-generated datasets 
provided by the Penn Machine Learning Benchmarks (PMLB)\cite{Olson2017PMLB,romano2021pmlb}. 
These datasets, generated using GAMETES\cite{Urbanowicz2012}, 
a software extensively employed in genetic studies 
for benchmarking machine learning algorithms. 
GAMETES simulates genetic data embodying feature interaction models 
of varying complexities and irrelevant features. 
Five of these datasets include 20 attributes 
generated using epistatic models with differing complexities. 
We also use another dataset 
containing 1000 attributes 
that serve as irrelevant features, 
thereby extending the 20-attributes datasets 
to configurations with 100 and 1000 attributes ({\tabSyntheticMetadata}). 
The resulting GAMETES datasets 
will be used to evaluate the effectiveness of our proposed method 
in the context of irrelevant features and epistasis.

\subsubsection{Colorectal cancer genotype data}
\label{sec:GWAS}

We will apply our proposed method 
to the colorectal cancer genotype dataset 
from transdisciplinary (CORECT) consortium~\cite{ScOt15}. 
Genotyping was conducted using a custom Affymetrix genome-wide platform 
on two physical genotyping chips, 
with a total of 696 samples (200 colorectal cancer cases and 496 controls) genotyped using the first chip and 
656 cases genotyped using the second chip. 
The data processing procedure consists of three parts: 
pre-imputation processing, imputation, and post-imputation processing.

We performed quality control using PLINK~\cite{ScOt15}. 
The pre-imputation process removed samples 
with a genotyping call rate less than 95\%, 
sex labeling not consistent with the chromosome, and 
sample heterozygosity not within three standard deviations from the mean. 
It also removed SNPs with a minor allele frequency less than 1\%. 
The two cohorts genotyped using different chips are merged and 
prepared to meet the requirements of the Michigan Imputation Server (MIS)~\cite{das2016next} 
following the guidelines provided by its official tutorial\footnote{https://imputationserver.readthedocs.io/en/latest/prepare-your-data/}. 
During the imputation step, the Michigan Imputation Server use 
the eagle phasing algorithm~\cite{loh2016reference}, the hg19 reference panel, and the option of mixed population to accommodate the multi-racial population structure. 
The post-imputation process excludes low-quality SNPs 
based on the imputation R2 of minimac3 ($\text{R2}>0.3$)~\cite{das2016next}. 
We extract all SNPs of the dataset used for MIS submission based on chromosome position and 
perform minor allele frequency ($>0.01$) and 
linkage disequilibrium filtering ($\text{r2}=0.2$). 
We will also remove first-degree relatives based on IBD. 
Samples in the second dataset were removed 
if the PI\_HAT value was above 0.5 with any samples in the first dataset.

Finally, a total of 197,497 SNPs and 1137 individuals are selected for subsequent analysis. 
Of which, 198 cases and 491 healthy controls are from the first dataset and 
448 cases are from the second.

\subsubsection{GEO datasets}
\label{sec:GEO}

In addition to synthetic data and genotype data, 
we extend the application of our proposed method to gene expression data. 
Specifically, we have chosen 13 gene expression datasets 
from CuMiDa~\cite{Feltes2019}, 
a repository that offers publicly accessible datasets 
suitable for machine learning analysis. 
Our selection criteria for these datasets are 
that each must contain more than 100 observations and a binary target label ({\tabSummaryGEO}).

\section{Results}

In this study, 
synthetic data is used to demonstrate 
the superiority of our proposed methodology 
in dealing with irrelevant features 
and feature interactions. 
We also apply our proposed methodology 
to genotype data and GEO datasets, 
in order to evaluate the prediction complexity 
posed by real micro-array datasets.

\subsection{Evaluation of conventional data complexity metrics using synthetic datasets}
The effectiveness of 
existing dataset complexity metrics 
for high-dimensional datasets 
remains underexplored. 
Many current complexity metrics 
depend on procedures 
which can be compromised 
by the existence of irrelevant features and epistasis, 
thus undermining their efficacy.

In this section, 
we first employ five synthetic datasets 
(referred to as G6, G7, G8, G9, and G10 in {\tabSyntheticMetadata}), 
provided by PMLB~\cite{Olson2017PMLB,romano2021pmlb}, 
to conduct complexity analysis. 
Each synthetic dataset comprises two balanced classes 
and every observation in the dataset 
is described using 20 features. 
The generation of these synthetic datasets 
is governed by several parameters, 
such as the degree of non-linear feature interaction, just name a few. 
These parameters describe the extent of correlation 
between the epistatic features and the class outcome.

To investigate the impact of irrelevant features,
we have also crafted three variations for each synthetic dataset. 
The first variation 
focuses on the complexity of the dataset sans irrelevant features 
(denoted as G1, G2, G3, G4, and G5 in {\tabSyntheticMetadata}). 
The second variation 
focuses on the scenario 
wherein the dataset encompasses a greater number of irrelevant features 
(labeled as G11, G12, G13, G14, and G15 in {\tabSyntheticMetadata}).
The third variation concentrates on the scenario 
when the dataset includes an excessive quantity of irrelevant features 
(referred to as G16, G17, G18, G19, and G20 in {\tabSyntheticMetadata}). 
Irrelevant features from a synthetic dataset of 1000 features on PMLB 
are used to create these variations. 
These variations are labelled 
"Low", "Median", and "High" in {\tabSyntheticMetadata}, 
and the original synthetic datasets labeled as "Normal".

Our complexity analysis, 
conducted using the Python library {\tt problexity}~\cite{komorniczak2023problexity}, 
shows that 
the effectiveness of multiple complexity metrics 
are impacted by the quantity of irrelevant features ({\tabSyntheticComplexity}). 
The complexity metrics 
can be encapsulated in a single scalar measure 
denoted as the \textit{score} in {\tabSyntheticComplexity}. 
A surge in the number of irrelevant features 
reduces the average \textit{score} from 0.576 (Low) to 0.513 (High).
Further, the average values of individual metrics 
are also affected by the increase of irrelevant features. 
A substantial decrease of \textit{f1v} (from 0.995 to 0.131),
\textit{f2} (from 1.0 to 0.0), 
and a slight decrease of the mean \textit{f4} (from 1.0 to 0.953) 
are observed as a result of the increase in irrelevant features. 
For linearity-based measures, 
significant decreases 
are observed for all three metrics 
when the quantity of irrelevant features 
is set to High. 
For neighborhood-based measures, 
\textit{n1} decreases from 0.305 (Low) to 0.248 (High), 
\textit{n4} decreases from 0.464 (Low) to 0.151 (High), 
and \textit{t1} increases from 0.935 (Low) to 1.0 (High). 
For network-based measures, 
minor increases of \textit{density} (from 0.855 to 1.0) 
and \textit{hubs} (from 0.754 to 0.996) are observed. For dimensionality-based measures, significant increases in \textit{t2} (from 0.002 to 0.625) and \textit{t3} (from 0.002 to 0.459) and a minor decrease of \textit{t4} (from 1.0 to 0.734) are observed.

The \textit{score} of the synthetic data 
that encapsulates 3-way epistasis is significantly 
lower than the average \textit{score} 
in the Low configuration. 
This indicates that 
an increase in the order of epistasis 
impacts the complexity metrics 
of synthetic datasets without 
irrelevant features (Low). 
The synthetic dataset with 3-way epistasis 
possesses lower linearity-based metrics 
(\textit{l1}: 0.287, \textit{l2}: 0.392, \textit{l3}: 0.401) 
compared to the average 
(\textit{l1}: 0.321, \textit{l2}: 0.467, \textit{l3}: 0.474).
For neighborhood-based metrics, 
\textit{n1} and \textit{n3} of the 3-way epistasis data 
are lower than those of the 2-way epistasis data (see {\tabSyntheticComplexity}). 
For network-based metrics, 
the \textit{density} (0.925) and \textit{hubs} (0.851) of the 3-way epistasis data 
are higher than the average (\textit{density}: 0.8554, \textit{hubs}: 0.754). 
However, these differences disappear 
with the increase of irrelevant features.

In summary, the existence of irrelevant features 
can impact the complexity measures, 
underscoring the necessity for complexity metrics 
that are robust to irrelevant features. 
Furthermore, in the realm of high-dimensional biological data analysis, 
the importance of feature interactions 
has been receiving increased attention~\cite{Liu2019}. 
However, existing complexity measures 
are ineffective in capturing the classification difficulty 
stemming from higher-order epistasis~\cite{Elizondo2012}. 
Although higher-order epistasis affects some metrics, 
these effects disappear when uncorrelated features are present ({\tabSyntheticComplexity}).

In the following sections, 
we elaborate on how our proposed method 
characterizes feature interactions and irrelevant features (Section~\ref{sec:synthtic_fitness_size-limit}). 
We also evaluate the synthetic datasets 
using our proposed complexity metrics 
to demonstrate the effectiveness of our approach (Section~\ref{sec:sythtic_proposed_complexity}).

\subsection{Evaluation of {\nameOfThisApproach} using synthetic datasets}
\label{sec:syn}

\begin{figure*}[h]
    \centering
    \includegraphics[width=\linewidth]{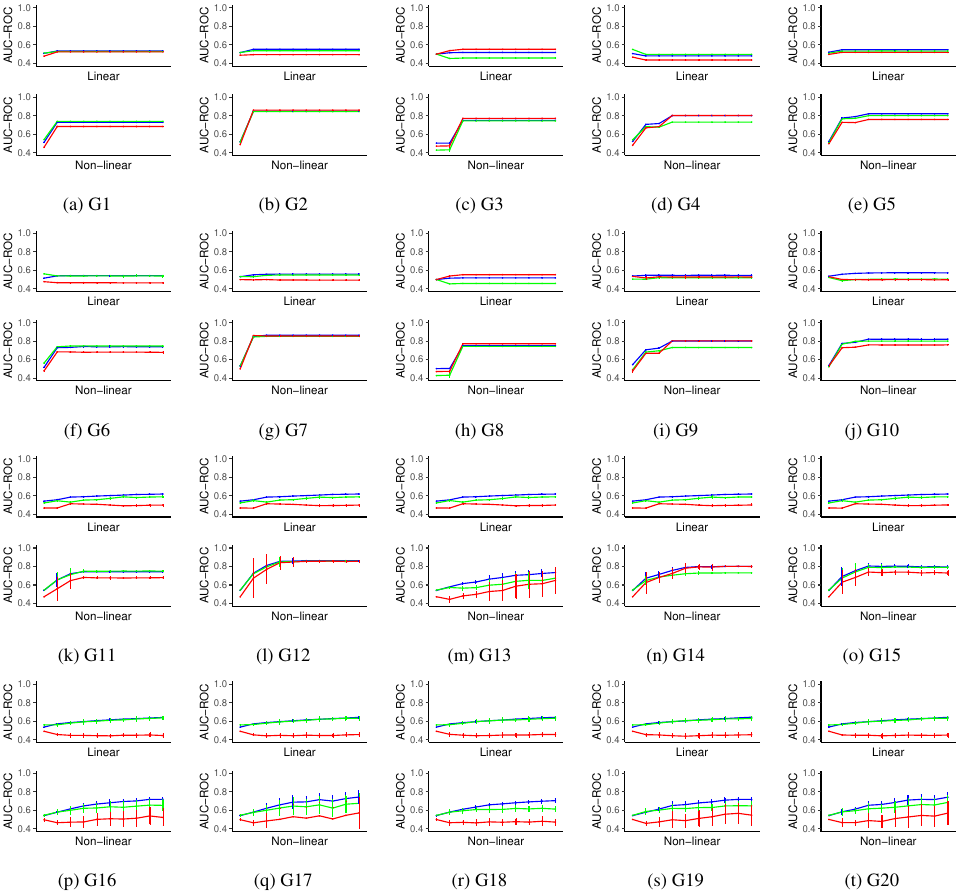}
    \caption{The depth plots for synthetic datasets. The \termSizeLimit~ranges from 1 to 10 and the averaged predictive performance on training (blue), testing (green) and validation (red) folds are represented on the y-axis. For each synthetic data, the depth plot based on linear and non-linear evaluation are provided.}
    \label{fig:synDepth}
\end{figure*}

\begin{table*}[h]
\centering
\caption{Complexity measures for lower dimentioinal synthetic data based on non-linear depth plot.}
\label{tab:SynComp-lo}
\begin{tabular}{ll||llllll|llllll}
\hline
            &  & \multicolumn{5}{l}{Low}  & $\overline{x}$ & \multicolumn{5}{l}{Normal} & $\overline{x}$   \\ \cline{3-7} \cline{9-13}
           & Data & G01 & G02 & G03 & G04 & G05 &   & G06 & G07 & G08 & G09 & G10 &  \\ \hline
Train      & 90\%    & 2  & 2  & 3  & 4  & 2  & 2.6 & 2  & 2  & 4  & 3  & 2   & 2.6 \\
           & 95\%    & 2  & 2  & 3  & 4  & 3  & 2.8 & 2  & 2  & 4  & 4  & 3   & 3   \\
           & 99\%    & 2  & 2  & 3  & 4  & 4  & 3   & 4  & 3  & 5  & 4  & 4   & 4   \\
           & 99.50\% & 2  & 2  & 3  & 4  & 4  & 3   & 4  & 3  & 5  & 4  & 4   & 4   \\
           & 100\%   & 2  & 2  & 3  & 4  & 4  & 3   & 4  & 6  & 10 & 5  & 4   & 5.8 \\
           & elbow & 2  & 2  & 3  & 4  & 2  & 2.6 & 2  & 2  & 4  & 4  & 4   & 3.2 \\ \hline
Test       & 90\%    & 2  & 2  & 3  & 2  & 2  & 2.2 & 2  & 2  & 4  & 2  & 2   & 2.4 \\
           & 95\%    & 2  & 2  & 3  & 4  & 2  & 2.6 & 2  & 2  & 4  & 3  & 2   & 2.6 \\
           & 99\%    & 2  & 2  & 3  & 4  & 4  & 3   & 3  & 2  & 5  & 4  & 3   & 3.4 \\
           & 99.50\% & 2  & 2  & 3  & 4  & 4  & 3   & 3  & 3  & 6  & 4  & 3   & 3.8 \\
           & 100\%   & 2  & 2  & 3  & 4  & 4  & 3   & 4  & 6  & 7  & 5  & 4   & 5.2 \\
           & elbow & 2  & 2  & 3  & 2  & 2  & 2.2 & 2  & 2  & 1  & 4  & 3   & 2.4 \\ \hline
Validation & 90\%    & 2  & 2  & 3  & 4  & 2  & 2.6 & 2  & 2  & 4  & 4  & 2   & 2.8 \\
           & 95\%    & 2  & 2  & 3  & 4  & 2  & 2.6 & 2  & 2  & 4  & 4  & 2   & 2.8 \\
           & 99\%    & 2  & 2  & 3  & 4  & 4  & 3   & 2  & 2  & 5  & 4  & 4   & 3.4 \\
           & 99.50\% & 2  & 2  & 3  & 4  & 4  & 3   & 2  & 2  & 6  & 4  & 4   & 3.6 \\
           & 100\%   & 2  & 2  & 3  & 4  & 4  & 3   & 2  & 2  & 6  & 5  & 4   & 3.8 \\
           & elbow & 2  & 2  & 3  & 4  & 2  & 2.6 & 2  & 2  & 1  & 4  & 2   & 2.2 \\ \hline
\end{tabular}
\end{table*}

\begin{table*}[t]
\centering
\caption{Complexity measures for higher dimentioinal synthetic data based on non-linear depth plot.}
\label{tab:SynComp-hi}
\begin{tabular}{ll||llllll|llllll}
\hline
            &  & \multicolumn{5}{l}{Median}  & $\overline{x}$ & \multicolumn{5}{l}{High} & $\overline{x}$   \\ \cline{3-7} \cline{9-13}
           & Data & G11 & G12 & G13 & G14 & G15 & & G16 & G17 & G18 & G19 & G20 &   \\ \hline
Train      & 90\%    & 3   & 3   & 5   & 4   & 3   & 3.6    & 4   & 5   & 4   & 4   & 6   & 4.6 \\
           & 95\%    & 3   & 4   & 7   & 5   & 4   & 4.6    & 7   & 7   & 6   & 7   & 7   & 6.8 \\
           & 99\%    & 4   & 4   & 10  & 6   & 4   & 5.6    & 9   & 10  & 9   & 9   & 10  & 9.4 \\
           & 99.50\% & 4   & 5   & 10  & 6   & 4   & 5.8    & 9   & 10  & 10  & 9   & 10  & 9.6 \\
           & 100\%   & 4   & 6   & 10  & 10  & 4   & 6.8    & 9   & 10  & 10  & 9   & 10  & 9.6 \\
           & elbow & 3   & 4   & 7   & 5   & 4   & 4.6    & 5   & 7   & 5   & 4   & 7   & 5.6 \\ \hline
Test       & 90\%    & 3   & 3   & 6   & 3   & 3   & 3.6    & 3   & 4   & 2   & 2   & 5   & 3.2 \\ 
           & 95\%    & 4   & 4   & 8   & 4   & 4   & 4.8    & 5   & 5   & 3   & 4   & 8   & 5   \\
           & 99\%    & 4   & 4   & 10  & 6   & 4   & 5.6    & 9   & 10  & 7   & 8   & 10  & 8.8 \\
           & 99.50\% & 4   & 6   & 10  & 6   & 4   & 6      & 9   & 10  & 7   & 8   & 10  & 8.8 \\
           & 100\%   & 9   & 9   & 10  & 10  & 6   & 8.8    & 9   & 10  & 9   & 9   & 10  & 9.4 \\
           & elbow & 4   & 4   & 2   & 3   & 4   & 3.4    & 6   & 5   & 4   & 4   & 4   & 4.6 \\ \hline
Validation & 90\%    & 3   & 4   & 7   & 4   & 3   & 4.2    & 1   & 5   & 1   & 6   & 7   & 4   \\
           & 95\%    & 4   & 4   & 10  & 5   & 4   & 5.4    & 8   & 9   & 1   & 8   & 8   & 6.8 \\
           & 99\%    & 4   & 6   & 10  & 6   & 4   & 6      & 9   & 10  & 1   & 9   & 10  & 7.8 \\
           & 99.50\% & 4   & 6   & 10  & 8   & 4   & 6.4    & 9   & 10  & 1   & 9   & 10  & 7.8 \\
           & 100\%   & 4   & 8   & 10  & 9   & 4   & 7      & 9   & 10  & 1   & 9   & 10  & 7.8 \\
           & elbow & 4   & 4   & 1   & 5   & 4   & 3.6    & 1   & 1   & 1   & 1   & 1   & 1   \\ \hline
\end{tabular}
\end{table*}

\subsubsection{The effect of fitness evaluations and \termSizeLimit~parameter}
\label{sec:synthtic_fitness_size-limit}

Our proposed approach 
deviates from directly applying machine learning algorithms, 
like Support Vector Machine (SVM) and 1-Nearest Neighbor (1-NN), to the dataset. 
Instead, it is built upon a feature selection algorithm 
that utilizes GA and leverages fitness evaluation based on decision tree algorithm
to account for epistasis. 
This section presents the dataset preparation and parameter configuration for GA, 
following which, we illustrate the capabilities of different fitness evaluations 
in capturing epistatic features. 
The impact of the {\termSizeLimit} parameter 
on the feature selection results 
is also discussed.

The dataset is segmented into three parts: 
training (60\%), testing (20\%), and validation (20\%) folds. 
The GA algorithm is supplied with the training and testing folds, 
whereas the validation fold 
is employed to assess the performance of the evolved feature subsets 
produced by the GA runs. 
To validate a feature subset, 
a predictive algorithm equivalent to the one employed in fitness evaluation, 
is trained using the training fold. 
The quality of the evolved feature subset 
is determined by assessing 
the area under the receiver operating characteristic curve (AUC-ROC) of this algorithm 
on the training, testing, and validation folds.
The GA parameters are configured as follows: 
a mutation rate of 0.2, 
a crossover rate of 0.8, 
a tournament size of 6, 
a population size of 500, 
and a limit of 50 generations. 
The use of a lower mutation rate and higher crossover rate 
is the rule of thumb to optimize the feature selection result. 
This experiment investigates 
the effects of fitness measures using different classifiers
(i.e., logistic regression and decision tree) and {\termSizeLimit} values 
ranging from 1 to 10. 
We execute 50 feature selection runs for GAs 
with diverse {\termSizeLimit} configurations.

We present the predictive performance results of all evolved feature subsets 
using the depth plot as described in Section~\ref{sec:depthPlot}. 
{\figSyntheticDepthPlot} shows that for all synthetic datasets with epistatic interactions, 
the GA runs with decision tree algorithm as the fitness measure can capture the epistatic features as the {\termSizeLimit} increases. 
However, the performance of GA with logistic regression as the fitness measure 
does not improve with the increase of {\termSizeLimit}.

Our results demonstrate that 
the decision tree fitness evaluation 
is crucial for GA-based feature selection to capture epistatic features. 
On the other hand, the logistic regression fitness measure is 
not effective in identifying feature subsets 
containing epistatic features, 
as logistic regression is based on linear regression 
and cannot recognize feature interactions. 
Furthermore, we find that increasing the {\termSizeLimit} enhances 
the GA's ability to identify target epistatic features. 
This phenomenon can be interpreted from two perspectives. 
First, the {\termSizeLimit} value 
should be greater than the order of the target epistatic features, 
as incomplete epistatic features do not provide evolutionary advantages 
during the GA's search process. 
Second, a further increase in the {\termSizeLimit}  
boost the GA's overall search power, 
increasing the possibility of capturing epistatic features.

\subsubsection{The complexity of synthetic datasets based on {\termSizeLimit}}
\label{sec:sythtic_proposed_complexity}
The complexity measure, {\nameOfThisApproach}, 
leverages the slope of the predictive performance 
as a function of {\termSizeLimit}
to quantify the complexity of dataset.
This section begins by briefly outlining 
how to extract the complexity metrics 
from the depth plot. 
This is followed by a comparison of the difficulties 
of synthetic datasets 
with different orders of epistatic features and epistatic heterogeneity. 
Epistatic heterogeneity means 
there are more than one epistatic model in the data.
Datasets with higher-order epistatic interaction or more epistatic models (heterogeneity)
are considered as more challenging for constructing a predictive algorithm.

The complexity determination procedure 
proposed in Section~\ref{sec:complexity} 
aims to identify the smallest size-limit 
required to achieve a certain degree of maximum performance 
in a depth plot. 
This procedure is repeated 
on training, testing, and validation data. 
The depth plot derived from the feature selection 
based on logistic regression 
is excluded from this section 
as logistic regression is incapable of capturing epistatic features.

Given the same number of features (N=20), 
identifying 3-way epistasis is evidently more challenging than 
identifying 2-way epistasis. 
\tabSyntheticProposedComplexity~compare the complexity of synthetic datasets 
containing 3-way epistasis with those containing 2-way epistasis. 
For all training, testing, and validation predictive performances, 
the complexity metrics (90\%) for 3-way epistasis datasets (training: 4, testing: 4, and validation: 4) 
are higher than the metrics for 2-way epistasis datasets. 
We also observe that the heritability parameter of synthetic data 
influences the resulting validation complexity measure, 
making the validation predictive performance of dataset G6 
smaller than that of G7. 
Heritability estimates the amount of variation in a particular trait 
that can be attributed to genetic variables~\cite{Eichler2010}.
As for the datasets containing heterogeneous epistatic models, 
their predictive performances converge 
when the \termSizeLimit~reaches four, 
suggesting the minimum \termSizeLimit~to detect two 2-way epistasis is four. 
The use of the 90\% cut-off 
is not applicable here 
because the performance gain of the second epistasis model is overlooked by 90\% cut-off.

The identification of epistatic features 
becomes more challenging 
when more irrelevant features are present, 
which requires larger \termSizeLimit~values 
for GA-based feature selection 
to achieve a stable identification. 
We observe that 
the complexity metrics (90\%) for the 3-way epistasis dataset 
configured as Median (training: 5, testing: 6, and validation: 7) 
are larger than those of the Normal configuration. 
We also find that the complexity metrics (90\%) 
for the 2-way epistasis dataset (training: 3, testing: 3, and validation: 3) 
are lower than those of the 3-way dataset. 
So we consider the proposed complexity metric 
to be effective in the Median configuration (100 attributes). 
However, these metrics are not indicative 
when the number of irrelevant features 
exceeds the discovery capability of the GA. 
We observe that the feature selection algorithm 
suffers from overfitting on the 3-way data configured as High (1000 features), 
which causes the validation curve in the depth plot 
to not reflect the complexity of the data.

Our results demonstrate the effectiveness of the proposed complexity metrics 
in the context of epistasis. 
Determining the threshold for the minimum \termSizeLimit~is crucial 
for the precision of the resulting complexity metric. 
Our findings suggest that 
using a 90\% threshold may overlook one of the two heterogeneous epistasis models 
due to one epistatic model 
capturing a sufficient amount of association. 
The precision of the complexity measure 
should depend on the requirements of the specific application domain. 
Furthermore, an increased number of irrelevant features 
can significantly hamper the effectiveness of the GA-based feature selection algorithm 
due to the exponentially increased search space, 
resulting in the failure of GA and 
crippling the efficacy of the proposed metrics. 
We attribute this to the search power of GA, 
which is largely determined by the population size, 
being overwhelmed by the increasing search space of the data. 
Therefore, we emphasize that 
the power of GA should be determined with great care, and 
successful deployment of the proposed approach depends on 
allocating sufficient search power to GA.

\subsection{Evaluation of {\nameOfThisApproach} using CRC genotype dataset}
\label{sec:gwas}

\begin{figure*}[h]
    \centering
    \includegraphics[width=.8\linewidth]{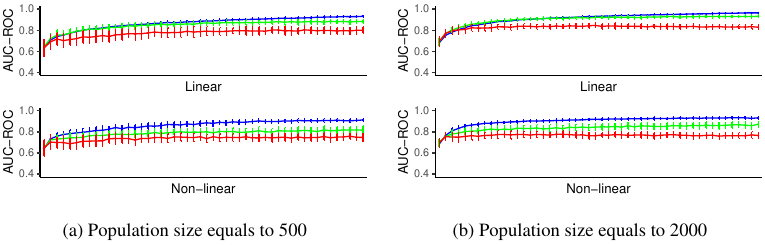}
    \caption{The depth plots for genotype dataset. The \termSizeLimit~ranges from 1 to 99 and the averaged predictive performance on training (blue), testing (green) and validation (red) folds are represented on the y-axis. For each population size configuration (500 and 2000), the depth plot based on linear and non-linear evaluation are provided.}
    \label{fig:GWASDepth}
\end{figure*}

\begin{table*}[t]
\centering
\caption{Complexity measures for genotype data based on depth plot.}
\label{tab:comp-GWAS}
\begin{tabular}{ll||lllll}
\hline
      & Population & 500                 &               &  & 2000                &               \\ \cline{3-4} \cline{6-7} 
      & Fitness & Linear & Non-linear &  & Linear & Non-linear \\ \hline
Train & 90\%            & 25                  & 21            &  & 17                  & 9             \\
      & 95\%            & 49                  & 39            &  & 39                  & 19            \\
      & 99\%            & 85                  & 75            &  & 83                  & 65            \\
      & 99.50\%         & 93                  & 91            &  & 87                  & 75            \\
      & 100\%           & 99                  & 99            &  & 99                  & 91            \\
      & elbow         & 27                  & 23            &  & 25                  & 19            \\ \hline
Test  & 90\%            & 13                  & 9             &  & 9                   & 7             \\
      & 95\%            & 27                  & 23            &  & 19                  & 25            \\
      & 99\%            & 71                  & 67            &  & 55                  & 83            \\
      & 99.50\%         & 73                  & 87            &  & 71                  & 93            \\
      & 100\%           & 89                  & 93            &  & 99                  & 99            \\
      & elbow         & 15                  & 15            &  & 25                  & 17            \\ \hline
Validation   & 90\%            & 13                  & 3             &  & 3                   & 3             \\
      & 95\%            & 27                  & 21            &  & 7                   & 3             \\
      & 99\%            & 71                  & 49            &  & 27                  & 19            \\
      & 99.50\%         & 71                  & 87            &  & 41                  & 19            \\
      & 100\%           & 81                  & 95            &  & 49                  & 39            \\
      & elbow         & 5                   & 3             &  & 11                  & 3  \\ \hline          
\end{tabular}
\end{table*}

The genotype data contain 
individual as well as epistatic feature associations. 
In addition to the complexity measure based on decision trees, 
we also provide a complexity measure 
based on logistic regression. 
We divide the genotype dataset into training (60\%), testing (20\%), and validation (20\%) sets and 
use the same GA parameters as shown in Section~\ref{sec:synthtic_fitness_size-limit}. 
We investigate every other \termSizeLimit~value 
ranging from 1 to 99 
by replicating feature selection runs 50 times. 
As the genotype data has approximately 200,000 features, 
we set the population size of GA to 2,000, 
ensuring that the initial coverage of the GA population (with \termSizeLimit~equals to 99) 
is equal to the number of features in the data. 
We consider that the population size not only affects the search capability of the GA 
but also determines the overall time complexity of the proposed method. 
Therefore, we also replicate the experiment 
with a population size of 500 
to examine the performance of the proposed complexity metrics 
on the real dataset with less search power of GA.

The results confirm that 
increasing the \termSizeLimit~improves 
the predictive performance of the feature subsets 
derived from GA (\figDepthPlotGWAS). 
The depth plot of the logistic regression algorithm 
shows that both training and testing performance 
increase as the \termSizeLimit~becomes larger. 
The highest validation performance (mean: 0.845) 
is achieved with a \termSizeLimit~of 49, 
indicating that a \termSizeLimit~larger than 49 may result in overfitting. 
The depth plot of the decision tree algorithm 
follows a similar pattern, 
with the highest validation performance (mean: 0.777) 
achieved with a \termSizeLimit~of 39. 
The complexity metrics for both logistic regression and decision tree 
are shown in \tabComplexityGWAS. 
We observe that the GA based on logistic regression and decision tree 
requires a minimum of 39 and 19 features, respectively, 
to achieve 95\% of the best training performance. 
When we address overfitting, 
the minimum \termSizeLimit~to achieve 95\% of the best validation performance 
decreases to 7 and 3, respectively.

The decrease in population size from 2,000 to 500 
slows down the convergence of predictive performance 
with respect to the \termSizeLimit~value. 
The best validation depth values (logistic regression: 81 and decision tree: 95) 
are larger than the results obtained from the GA 
with a population size of 2,000. 
In addition, the best averaged validation performances (logistic regression: 0.8032, decision tree: 0.7592) 
are not as good as the results 
obtained from GA with a population size of 2,000 (logistic regression: 0.845, decision tree: 0.777).

GA is a heuristic approach to feature selection. 
Our goal is to characterize the difficulty of the prediction problem 
for a given dataset using the slope of the predictive performance change. 
Therefore, the actual population size to be taken can be determined with flexibility 
and based on the computational resources that can be deployed. 
Nevertheless, when comparing the difficulties of different datasets, 
the population size for GA should be consistent across different datasets, 
in order to ensure that the resulting complexity values are comparable.

\subsection{Evaluation of {\nameOfThisApproach} using GEO datasets}
\label{sec:geo}

\begin{figure*}[h]
    \centering
    \includegraphics[width=\linewidth]{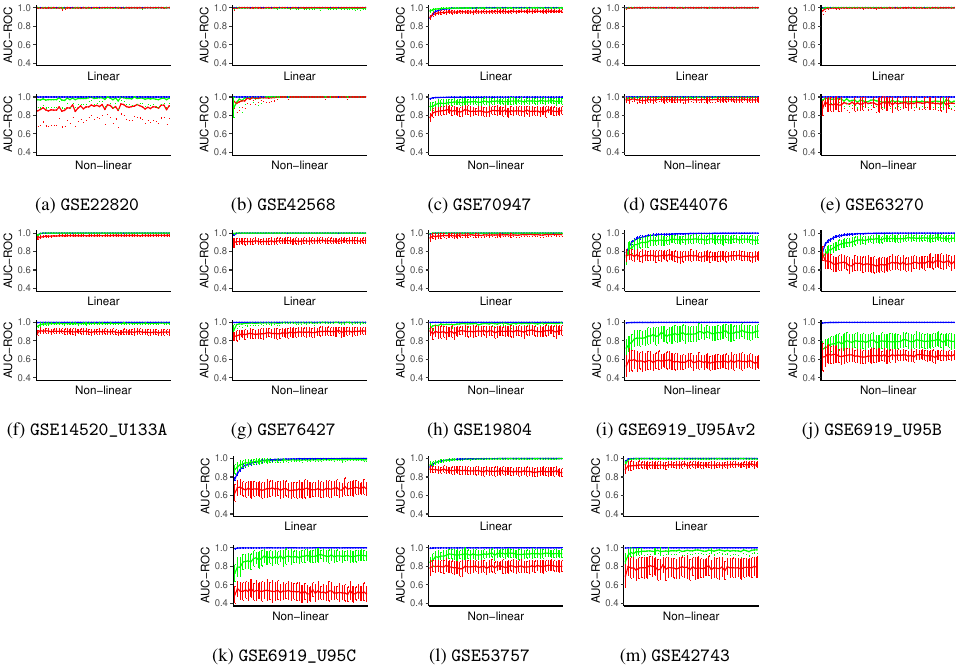}
    \caption{The depth plots for GEO datasets. The \termSizeLimit~ranges from 1 to 99 and the averaged predictive performance on training (blue), testing (green) and validation (red) folds are represented on the y-axis. For each GEO datasets, the depth plot based on linear and non-linear evaluation are provided.}
    \label{fig:GEODepth}
\end{figure*}

\begin{table*}[t]
\centering
\caption{Complexity measures for GEO datas based on the validation curve of depth plot.}
\label{tabcompGeo}
\begin{tabular}{ll||llllll}
\hline
Data            & Fitness    & 90\% & 95\% & 99\% & 99.50\% & 100\% & elbow \\ \hline
{\tt GSE22820}        & Linear     & 1    & 1    & 1    & 1       & 13    & 3     \\
        & Non-linear & 1    & 5    & 49   & 63      & 63    & 5     \\
{\tt GSE42568}        & Linear     & 1    & 1    & 1    & 3       & 9     & 9     \\
        & Non-linear & 1    & 1    & 15   & 33      & 35    & 1     \\
{\tt GSE70947}        & Linear     & 1    & 3    & 15   & 35      & 63    & 11    \\
        & Non-linear & 1    & 3    & 13   & 15      & 15    & 15    \\
{\tt GSE44076}        & Linear     & 1    & 1    & 1    & 3       & 63    & 3     \\
        & Non-linear & 1    & 1    & 1    & 5       & 89    & 1     \\
{\tt GSE63270}        & Linear     & 1    & 1    & 5    & 7       & 65    & 9     \\
        & Non-linear & 1    & 3    & 3    & 3       & 51    & 3     \\
{\tt GSE14520\_U133A} & Linear     & 1    & 1    & 7    & 13      & 93    & 13    \\
 & Non-linear & 1    & 1    & 3    & 3       & 3     & 3     \\
{\tt GSE76427}        & Linear     & 1    & 1    & 17   & 41      & 67    & 3     \\
        & Non-linear & 1    & 7    & 71   & 79      & 79    & 7     \\
{\tt GSE19804}        & Linear     & 1    & 1    & 23   & 31      & 93    & 3     \\
        & Non-linear & 1    & 1    & 41   & 77      & 91    & 1     \\
{\tt GSE6919\_U95Av2} & Linear     & 1    & 1    & 3    & 3       & 33    & 3     \\
 & Non-linear & 3    & 3    & 3    & 7       & 7     & 3     \\
{\tt GSE6919\_U95B}   & Linear     & 1    & 1    & 1    & 1       & 1     & 1     \\
   & Non-linear & 1    & 3    & 5    & 5       & 5     & 5     \\
{\tt GSE6919\_U95C}   & Linear     & 1    & 3    & 5    & 83      & 83    & 5     \\
   & Non-linear & 3    & 3    & 13   & 13      & 13    & 3     \\
{\tt GSE53757}        & Linear     & 1    & 1    & 1    & 1       & 11    & 1     \\
        & Non-linear & 1    & 1    & 13   & 13      & 91    & 3     \\
{\tt GSE42743}        & Linear     & 1    & 3    & 45   & 45      & 45    & 11    \\
        & Non-linear & 3    & 3    & 19   & 29      & 29    & 3     \\ \hline
\end{tabular}
\end{table*}

We also investigate the complexity of 13 gene-expression datasets available from CuMiDa~\cite{Feltes2019} (\tabSummaryGEO). 
As these datasets have a smaller number of features 
than the genotype dataset, 
we set the population size of the GA to 500 and 
keep the other parameters unchanged.

The results show that 
the overall pattern of the depth plot for all GEO datasets (\figDepthPlotGEO) 
is similar to the plot of the genotype dataset. 
However, the feature selection runs for three Prostate datasets ({\tt GSE6919\_U95B}, {\tt GSE6919\_U95Av2} and {\tt GSE6919\_U95C}) are unsuccessful 
due to overfitting is observed on the validation fold. 
For the remaining 10 datasets, 
the complexity measures (\tabComplexityGEO) based on the validation curves 
suggest that most datasets only need 3 features to achieve 95\% of the best validation performance. 
Any further improvement in validation performance 
requires increasing the number of features. 
The linear fitness-based validation complexities 
indicate that only a fraction of the datasets 
have a depth of 3 for achieving 95\% validation performance, 
while the rest have a difficulty of 1. 
If the complexity threshold is increased to 99\%, 
the most difficult data set changes to {\tt GSE19804}. 
The nonlinear fitness-based validation complexity 
indicates that the most difficult datasets are {\tt GSE76427} (95\%: 7) and {\tt GSE22820} (95\%: 5).

In summary, the complexity of the genotype dataset 
is higher than that of the GEO datasets. 
Due to the fact that feature subsets in the genotype data have a higher degree of overfitting, the validation complexity of GEO datasets 
converges faster and has lower a complexity than the genotype dataset. 
Nevertheless, the training complexity of
the GEO data is still significantly lower than that of the genotype data. 
It is not surprising that 
GEO data is less complex than genotype data, 
and our results elucidate this fact based on \nameOfThisApproach.

\section{Conclusion}

\label{sec:conclusion}
We proposed a  robust complexity metric--{\nameOfThisApproach},
for biomedical data that may contain 
irrelevant features and epistasis. 
The limitation of existing data complexity metrics lies in that
they work well for low-dimensional synthetic data
but are challenged in the presence of increasing irrelevant features and feature interactions, 
as shown in {\tabSyntheticComplexity}.

The main contribution of our approach is that 
{\nameOfThisApproach} integrates a well-established feature selection method 
and gives it a new meaning.
Our methodology includes two components, 
1) the implementation of a feature selection algorithm, 
which diminishes the impact of irrelevant features,
and 2) the deployment of a decision tree predictive algorithm. 
This allows the proposed complexity measure 
to detect data complexity 
arising from feature interactions. 
The results obtained from this approach 
suggest that by controlling the value of {\termSizeLimit}, 
our proposed complexity measure 
is capable of distinguishing the complexity of synthetic data 
resulted by different orders of epistatic interactions ({\figSyntheticDepthPlot} and {\tabSyntheticProposedComplexity}).

Nevertheless, our proposed methodology {\nameOfThisApproach} 
works less effective on synthetic 3-way epistasis data with 1000 features. 
There could be two major reasons for this.
Firstly, the expansion of the feature set 
considerably enlarges the search space 
for the feature selection algorithm, 
thereby exceeding its search capacity. 
Secondly, the increasing presence of irrelevant features 
could increase false positives for feature selection. 
Consequently, the selected feature subsets 
may demonstrate fitness values 
comparable to those of the target features 
on the training dataset 
but fail to replicate their performance 
on the validation dataset.

\subsection{The power of feature selection}
\label{sec:conclusion-power}

The effectiveness of our feature selection genetic algorithm 
depends on its parameters, 
specifically the population size 
and the {\termSizeLimit}. 
The population size determines the quantity of feature subsets 
that the algorithm evaluates. 
An increase in population size 
enhances the algorithm's capacity 
to discover new feature subsets.
However, this also increases 
the computational burden for the algorithm. 
Our complexity analysis conducted on 
simulated and genotype data indicates 
that the expansion of the search space
and a decrease in population size, 
distorts the depth plot's curve, 
resulting in a higher complexity score. 
Moreover, our assessment of synthetic data with 3-way epistasis and 1000 features 
suggests that larger population size 
can not guarantee the identification of epistatic feature. 
As such, we consider the proposed method augments the resilience of the metrics to irrelevant features.

In contrast, altering the {\termSizeLimit} 
enhances the fundamental discovery capability of the algorithm.
The resulting depth plot 
is a graphical representation 
illustrating the predictive performance of the chosen feature subsets 
as a function of the size of selected feature subsets. 
Utilizing a larger {\termSizeLimit }
allows the algorithm 
to uncover larger feature subsets 
containing more relevant features. 
Nevertheless, a larger {\termSizeLimit} 
can increase the risk of overfitting. 
We have noted that overfitting 
is a prevalent issue in real data ({\figDepthPlotGWAS} and {\figDepthPlotGEO}). 
Consequently, it is essential 
to consider the validation curve in the depth plot.

\subsection{Complexity measures for micro-array data}
Unlike synthetic data, 
real data exhibits both linear and non-linear associations between features and target labels. 
To accommodate these characteristics, 
we employ two complementary fitness measures 
for evaluating data complexity. 
Based on the derived complexity metrics, 
we subsequently offer guidelines 
for interpreting the complexity outcomes.

In the context of genotype data, 
the complexity deduced from the linear depth curve 
appears to surpass that of the non-linear curve, 
as illustrated in ({\tabComplexityGWAS}). 
This disparity may stem from the GA's 
restricted search capabilities. 
Our simulation study suggests that 
the GA-based feature selection method 
may struggle to identify 3-way epistasis in synthetic data with 1000 features (G18 in {\figSyntheticDepthPlot}). 
Furthermore, we observe that 
decreasing the parameter population size 
results in a flatter depth curve in the plot, 
thereby skewing the complexity estimation. 
Nevertheless, a reduction in computational power (need for a greater population size)
does not influence the relative difference 
between the linear and non-linear complexity 
in terms of validation predictive performance. 
In other words, the relative complexity of the data 
remains invariant to the search power.

According to our complexity metrics, 
the GEO data seems less complex compared to genotype data.
When employing equivalent search power (population size of 500), 
we find that both the linear and non-linear difficulties (validation at 99\%) of genotype data exceed those of most GEO datasets. 
An exception is the non-linear validation difficulty of {\tt GSE76427}, 
although its 95\% validation difficulty (7) is lower than that of the genotype data (21). 
Based on the validation curve of the majority of GEO datasets, 
achieving a 90\% predictive performance only need the use of just one feature. 
This implies that a feature selection method based on Differentially Expressed Genes (DEG) is appropriate for most GEO datasets.

Our proposed complexity measure 
enables us to estimate the difficulty of a prediction problem for a dataset from both linear and non-linear perspectives. 
This helps with crafting a prediction model that aligns with the data's specific properties. 
Moreover, the depth curve in the plot can be utilized to determine the optimal number of features for the feature selection method. 
Selecting a smaller set of features enhances interpretability and reduces the risk of overfitting.

\subsection{Limitations and future work}
The feature selection algorithm 
is important for the efficacy of our proposed data complexity measure. 
In our feature selection genetic algorithm,
its parameter population size is a good indicator of its search effort.
The computational overhead resulted by increased population size can be
a limitation of our approach.

Exploring more efficient feature selection methods 
can be a promising next-step direction. 
A more efficient search strategy than GA 
may reduce the computational cost of the complexity measure. 
For high-dimensional data,
the search space increase exponentially with the number of features.
So increasing population size of the GA may not be always effective. 
We use GA as the search strategy 
also because its working processes, including \termSizeLimit, mutation and crossover, 
are highly interpretable and controllable.

\section*{Supplementary materials}
\begin{itemize}
    \item Table S1: The metadata of synthetic datasets.
    \item Table S2: The complexity measures for synthetic datasets based on {\tt problixity}.
    \item Table S3: The metadata for GEO datasets.
    \item Table S4: The complexity measures for GEO datasets based on {\tt problixity}.
\end{itemize}

\section*{Data availability}
The simulation datasets can be accessed through the Penn Machine Learning Benchmarks~\cite{Olson2017PMLB,romano2021pmlb}. 
The genotype data can be accessed through the colorectal cancer transdisciplinary (CORECT) consortium~\cite{ScOt15}.
The GEO datasets can be accessed through the Curated Machine Learning Datasets (CuMiDa)~\cite{Feltes2019}.

\section*{Code Availability Statement}

The code that supports the findings of this study is openly accessible. It can be found on the GitHub platform at the following link\footnote{https://github.com/shazhendong/DataComplexityAnalysisUsingGA}.

\section*{Acknowledgments}

We are grateful to Digital Research Alliance of Canada and
Wireless Networking and Mobile Computing Laboratory
for providing computing infrastructures.
We are also grateful to Matthew Vandergrift for providing feedback on the manuscript. 

\bibliographystyle{IEEEtran}
\bibliography{references}

\begin{thebibliography}{10}
\providecommand{\url}[1]{#1}
\csname url@samestyle\endcsname
\providecommand{\newblock}{\relax}
\providecommand{\bibinfo}[2]{#2}
\providecommand{\BIBentrySTDinterwordspacing}{\spaceskip=0pt\relax}
\providecommand{\BIBentryALTinterwordstretchfactor}{4}
\providecommand{\BIBentryALTinterwordspacing}{\spaceskip=\fontdimen2\font plus
\BIBentryALTinterwordstretchfactor\fontdimen3\font minus
  \fontdimen4\font\relax}
\providecommand{\BIBforeignlanguage}[2]{{%
\expandafter\ifx\csname l@#1\endcsname\relax
\typeout{** WARNING: IEEEtran.bst: No hyphenation pattern has been}%
\typeout{** loaded for the language `#1'. Using the pattern for}%
\typeout{** the default language instead.}%
\else
\language=\csname l@#1\endcsname
\fi
#2}}
\providecommand{\BIBdecl}{\relax}
\BIBdecl

\bibitem{Libbrecht2015}
M.~W. Libbrecht and W.~S. Noble, ``Machine learning applications in genetics
  and genomics,'' \emph{Nature Reviews Genetics}, vol.~16, no.~6, pp. 321--332,
  Jun 2015.

\bibitem{Nicholls2020}
H.~L. Nicholls, C.~R. John, D.~S. Watson, P.~B. Munroe, M.~R. Barnes, and C.~P.
  Cabrera, ``Reaching the end-game for gwas: Machine learning approaches for
  the prioritization of complex disease loci,'' \emph{Frontiers in Genetics},
  vol.~11, 2020.

\bibitem{Bauer2014}
D.~C. Bauer, C.~Gaff, M.~E. Dinger, M.~Caramins, F.~A. Buske, M.~Fenech,
  D.~Hansen, and L.~Cobiac, ``Genomics and personalised whole-of-life
  healthcare,'' \emph{Trends in Molecular Medicine}, vol.~20, no.~9, pp.
  479--486, 2014.

\bibitem{Liu2019}
X.~Liu, Y.~I. Li, and J.~K. Pritchard, ``Trans effects on gene expression can
  drive omnigenic inheritance,'' \emph{Cell}, vol. 177, no.~4, pp.
  1022--1034.e6, May 2019.

\bibitem{ritchie2001multifactor}
M.~D. Ritchie, L.~W. Hahn, N.~Roodi, L.~R. Bailey, W.~D. Dupont, F.~F. Parl,
  and J.~H. Moore, ``Multifactor-dimensionality reduction reveals high-order
  interactions among estrogen-metabolism genes in sporadic breast cancer,''
  \emph{The American Journal of Human Genetics}, vol.~69, no.~1, pp. 138--147,
  2001.

\bibitem{Ho2002}
T.~K. Ho and M.~Basu, ``Complexity measures of supervised classification
  problems,'' \emph{IEEE Transactions on Pattern Analysis and Machine
  Intelligence}, vol.~24, no.~3, pp. 289--300, 2002.

\bibitem{Komorniczak2023}
J.~Komorniczak and P.~Ksieniewicz, ``problexity—an open-source python library
  for supervised learning problem complexity assessment,''
  \emph{Neurocomputing}, vol. 521, pp. 126--136, 2023.

\bibitem{vanschoren2018meta}
J.~Vanschoren, ``Meta-learning: A survey,'' \emph{arXiv preprint
  arXiv:1810.03548}, 2018.

\bibitem{Meskhi2021}
M.~M. Meskhi, A.~Rivolli, R.~G. Mantovani, and R.~Vilalta, ``Learning abstract
  task representations,'' in \emph{AAAI Workshop on Meta-Learning and MetaDL
  Challenge}, ser. Proceedings of Machine Learning Research, I.~Guyon, J.~N.
  van Rijn, S.~Treguer, and J.~Vanschoren, Eds., vol. 140.\hskip 1em plus 0.5em
  minus 0.4em\relax PMLR, 09 Feb 2021, pp. 127--137.

\bibitem{Ho2022}
T.~Kam~Ho, ``Complexity of representations in deep learning,'' in \emph{2022
  26th International Conference on Pattern Recognition (ICPR)}, 2022, pp.
  2657--2663.

\bibitem{Konuk_2019_ICCV}
E.~Konuk and K.~Smith, ``An empirical study of the relation between network
  architecture and complexity,'' in \emph{Proceedings of the IEEE/CVF
  International Conference on Computer Vision (ICCV) Workshops}, Oct 2019.

\bibitem{moore2009}
J.~H. Moore and S.~M. Williams, ``Epistasis and its implications for personal
  genetics,'' \emph{The American Journal of Human Genetics}, vol.~85, no.~3,
  pp. 309--320, 2009.

\bibitem{orriols2010documentation}
A.~Orriols-Puig, N.~Macia, and T.~K. Ho, ``Documentation for the data
  complexity library in c++,'' \emph{Universitat Ramon Llull, La Salle}, vol.
  196, no. 1-40, p.~12, 2010.

\bibitem{HO1998}
T.~K. Ho and H.~S. Baird, ``Pattern classification with compact distribution
  maps,'' \emph{Computer Vision and Image Understanding}, vol.~70, no.~1, pp.
  101--110, 1998.

\bibitem{Smith1968}
F.~Smith, ``Pattern classifier design by linear programming,'' \emph{IEEE
  Transactions on Computers}, vol. C-17, no.~4, pp. 367--372, 1968.

\bibitem{Hoekstra1996}
A.~Hoekstra and R.~Duin, ``On the nonlinearity of pattern classifiers,'' in
  \emph{Proceedings of 13th International Conference on Pattern Recognition},
  vol.~4, 1996, pp. 271--275 vol.4.

\bibitem{Friedman1979}
J.~H. Friedman and L.~C. Rafsky, ``{Multivariate Generalizations of the
  Wald-Wolfowitz and Smirnov Two-Sample Tests},'' \emph{The Annals of
  Statistics}, vol.~7, no.~4, pp. 697 -- 717, 1979.

\bibitem{Cover1967}
T.~Cover and P.~Hart, ``Nearest neighbor pattern classification,'' \emph{IEEE
  Transactions on Information Theory}, vol.~13, no.~1, pp. 21--27, 1967.

\bibitem{Leyva2015}
E.~Leyva, A.~González, and R.~Pérez, ``A set of complexity measures designed
  for applying meta-learning to instance selection,'' \emph{IEEE Transactions
  on Knowledge and Data Engineering}, vol.~27, no.~2, pp. 354--367, 2015.

\bibitem{Gower1971}
J.~C. Gower, ``A general coefficient of similarity and some of its
  properties,'' \emph{Biometrics}, vol.~27, no.~4, pp. 857--871, 1971.

\bibitem{Lorena2019}
A.~C. Lorena, L.~P.~F. Garcia, J.~Lehmann, M.~C.~P. Souto, and T.~K. Ho, ``How
  complex is your classification problem? a survey on measuring classification
  complexity,'' \emph{ACM Comput. Surv.}, vol.~52, no.~5, sep 2019.

\bibitem{basu2006data}
M.~Basu and T.~K. Ho, \emph{Data complexity in pattern recognition}.\hskip 1em
  plus 0.5em minus 0.4em\relax Springer Science \& Business Media, 2006.

\bibitem{LORENA201233}
A.~C. Lorena, I.~G. Costa, N.~Spolaôr, and M.~C. {de Souto}, ``Analysis of
  complexity indices for classification problems: Cancer gene expression
  data,'' \emph{Neurocomputing}, vol.~75, no.~1, pp. 33--42, 2012, brazilian
  Symposium on Neural Networks (SBRN 2010) International Conference on Hybrid
  Artificial Intelligence Systems (HAIS 2010).

\bibitem{Tanwani2010}
A.~K. Tanwani and M.~Farooq, ``Classification potential vs. classification
  accuracy: A comprehensive study of evolutionary algorithms with biomedical
  datasets,'' in \emph{Learning Classifier Systems}, J.~Bacardit, W.~Browne,
  J.~Drugowitsch, E.~Bernad{\'o}-Mansilla, and M.~V. Butz, Eds.\hskip 1em plus
  0.5em minus 0.4em\relax Berlin, Heidelberg: Springer Berlin Heidelberg, 2010,
  pp. 127--144.

\bibitem{Lorena2018}
A.~C. Lorena, A.~I. Maciel, P.~B.~C. de~Miranda, I.~G. Costa, and R.~B.~C.
  Prud{\^e}ncio, ``Data complexity meta-features for regression problems,''
  \emph{Machine Learning}, vol. 107, no.~1, pp. 209--246, Jan 2018.

\bibitem{Anders2010}
S.~Anders and W.~Huber, ``Differential expression analysis for sequence count
  data,'' \emph{Nature Precedings}, Mar 2010.

\bibitem{siedlecki1993note}
W.~Siedlecki and J.~Sklansky, ``A note on genetic algorithms for large-scale
  feature selection,'' in \emph{Handbook of pattern recognition and computer
  vision}.\hskip 1em plus 0.5em minus 0.4em\relax World Scientific, 1993, pp.
  88--107.

\bibitem{al2017examining}
M.~Al-Rajab, J.~Lu, and Q.~Xu, ``Examining applying high performance genetic
  data feature selection and classification algorithms for colon cancer
  diagnosis,'' \emph{Computer Methods and Programs in Biomedicine}, vol. 146,
  pp. 11--24, 2017.

\bibitem{swerhun2020summary}
M.~Swerhun, J.~Foley, B.~Massop, and V.~Mago, ``A summary of the prevalence of
  genetic algorithms in bioinformatics from 2015 onwards,'' \emph{arXiv
  preprint arXiv:2008.09017}, 2020.

\bibitem{sayed2019nested}
S.~Sayed, M.~Nassef, A.~Badr, and I.~Farag, ``A nested genetic algorithm for
  feature selection in high-dimensional cancer microarray datasets,''
  \emph{Expert Systems with Applications}, vol. 121, pp. 233--243, 2019.

\bibitem{garcia2020unsupervised}
P.~Garc{\'\i}a-D{\'\i}az, I.~S{\'a}nchez-Berriel, J.~A. Mart{\'\i}nez-Rojas,
  and A.~M. Diez-Pascual, ``Unsupervised feature selection algorithm for
  multiclass cancer classification of gene expression rna-seq data,''
  \emph{Genomics}, vol. 112, no.~2, pp. 1916--1925, 2020.

\bibitem{leardi1992genetic}
R.~Leardi, R.~Boggia, and M.~Terrile, ``Genetic algorithms as a strategy for
  feature selection,'' \emph{Journal of Chemometrics}, vol.~6, no.~5, pp.
  267--281, 1992.

\bibitem{Sha2021}
Z.~Sha, T.~Hu, and Y.~Chen, ``Feature selection for polygenic risk scores using
  genetic algorithm and network science,'' in \emph{2021 IEEE Congress on
  Evolutionary Computation (CEC)}, 2021, pp. 802--808.

\bibitem{da2011improving}
S.~F. Da~Silva, M.~X. Ribeiro, J.~d. E.~B. Neto, C.~Traina-Jr, and A.~J.
  Traina, ``Improving the ranking quality of medical image retrieval using a
  genetic feature selection method,'' \emph{Decision Support Systems}, vol.~51,
  no.~4, pp. 810--820, 2011.

\bibitem{canuto2012genetic}
A.~M. Canuto and D.~S. Nascimento, ``A genetic-based approach to features
  selection for ensembles using a hybrid and adaptive fitness function,'' in
  \emph{The 2012 International Joint Conference on Neural Networks
  (IJCNN)}.\hskip 1em plus 0.5em minus 0.4em\relax IEEE, 2012, pp. 1--8.

\bibitem{sousa2013email}
P.~Sousa, P.~Cortez, R.~Vaz, M.~Rocha, and M.~Rio, ``Email spam detection: A
  symbiotic feature selection approach fostered by evolutionary computation,''
  \emph{International Journal of Information Technology \& Decision Making},
  vol.~12, no.~04, pp. 863--884, 2013.

\bibitem{seo2014feature}
J.-H. Seo, Y.~H. Lee, and Y.-H. Kim, ``Feature selection for very short-term
  heavy rainfall prediction using evolutionary computation,'' \emph{Advances in
  Meteorology}, vol. 2014, 2014.

\bibitem{winkler2011identification}
S.~M. Winkler, M.~Affenzeller, W.~Jacak, and H.~Stekel, ``Identification of
  cancer diagnosis estimation models using evolutionary algorithms: a case
  study for breast cancer, melanoma, and cancer in the respiratory system,'' in
  \emph{Proceedings of the 13th Annual Conference Companion on Genetic and
  Evolutionary Computation}, 2011, pp. 503--510.

\bibitem{souza2011co}
F.~Souza, T.~Matias, and R.~Ara{\'o}jo, ``Co-evolutionary genetic multilayer
  perceptron for feature selection and model design,'' in
  \emph{ETFA2011}.\hskip 1em plus 0.5em minus 0.4em\relax IEEE, 2011, pp. 1--7.

\bibitem{oreski2014genetic}
S.~Oreski and G.~Oreski, ``Genetic algorithm-based heuristic for feature
  selection in credit risk assessment,'' \emph{Expert Systems with
  Applications}, vol.~41, no.~4, pp. 2052--2064, 2014.

\bibitem{breiman2001random}
L.~Breiman, ``Random forests,'' \emph{Machine learning}, vol.~45, no.~1, pp.
  5--32, 2001.

\bibitem{dorani2018ensemble}
F.~Dorani, T.~Hu, M.~O. Woods, and G.~Zhai, ``Ensemble learning for detecting
  gene-gene interactions in colorectal cancer,'' \emph{PeerJ}, vol.~6, p.
  e5854, 2018.

\bibitem{PHMA13}
Q.~Pan, T.~Hu, J.~D. Malley, A.~S. Andrew, M.~R. Karagas, and J.~H. Moore,
  ``Supervising random forest using attribute interaction networks,'' in
  \emph{European Conference on Evolutionary Computation, Machine Learning and
  Data Mining in Bioinformatics}.\hskip 1em plus 0.5em minus 0.4em\relax
  Springer, 2013, pp. 104--116.

\bibitem{MoAW10}
J.~H. Moore, F.~W. Asselbergs, and S.~M. Williams, ``Bioinformatics challenges
  for genome-wide association studies,'' \emph{Bioinformatics}, vol.~26, no.~4,
  pp. 445--455, 2010.

\bibitem{Mak2017}
T.~S.~H. Mak, R.~M. Porsch, S.~W. Choi, X.~Zhou, and P.~C. Sham, ``Polygenic
  scores via penalized regression on summary statistics,'' \emph{Genetic
  Epidemiology}, vol.~41, no.~6, pp. 469--480, 2017.

\bibitem{wang2010antepiseeker}
Y.~Wang, X.~Liu, K.~Robbins, and R.~Rekaya, ``Antepiseeker: detecting epistatic
  interactions for case-control studies using a two-stage ant colony
  optimization algorithm,'' \emph{BMC research notes}, vol.~3, no.~1, pp. 1--8,
  2010.

\bibitem{storn1997differential}
R.~Storn and K.~Price, ``Differential evolution--a simple and efficient
  heuristic for global optimization over continuous spaces,'' \emph{Journal of
  global optimization}, vol.~11, no.~4, pp. 341--359, 1997.

\bibitem{cao2019hissi}
X.~Cao, J.~Liu, M.~Guo, and J.~Wang, ``Hissi: high-order snp-snp interactions
  detection based on efficient significant pattern and differential
  evolution,'' \emph{BMC medical genomics}, vol.~12, no.~7, pp. 1--12, 2019.

\bibitem{guan2021differential}
B.~Guan, Y.~Zhao, Y.~Yin, and Y.~Li, ``A differential evolution based feature
  combination selection algorithm for high-dimensional data,''
  \emph{Information Sciences}, vol. 547, pp. 870--886, 2021.

\bibitem{Holland1992}
J.~H. Holland, ``Genetic algorithms,'' \emph{Scientific American}, vol. 267,
  no.~1, pp. 66--73, 1992.

\bibitem{miller1995genetic}
B.~L. Miller, D.~E. Goldberg \emph{et~al.}, ``Genetic algorithms, tournament
  selection, and the effects of noise,'' \emph{Complex systems}, vol.~9, no.~3,
  pp. 193--212, 1995.

\bibitem{fisher_1919}
R.~A. Fisher, ``Xv.—the correlation between relatives on the supposition of
  mendelian inheritance.'' \emph{Earth and Environmental Science Transactions
  of The Royal Society of Edinburgh}, vol.~52, no.~2, p. 399–433, 1919.

\bibitem{Zhou_2022}
J.~Zhou, M.~S. Wong, W.-C. Chen, A.~R. Krainer, J.~B. Kinney, and D.~M.
  McCandlish, ``Higher-order epistasis and phenotypic prediction,''
  \emph{Proceedings of the National Academy of Sciences}, vol. 119, no.~39, p.
  e2204233119, 2022.

\bibitem{Cox1958}
D.~R. Cox, ``The regression analysis of binary sequences,'' \emph{Journal of
  the Royal Statistical Society: Series B (Methodological)}, vol.~20, no.~2,
  pp. 215--232, 1958.

\bibitem{choi2020tutorial}
S.~W. Choi, T.~S.-H. Mak, and P.~F. O’Reilly, ``Tutorial: a guide to
  performing polygenic risk score analyses,'' \emph{Nature Protocols}, vol.~15,
  no.~9, pp. 2759--2772, 2020.

\bibitem{breiman2017classification}
L.~Breiman, \emph{Classification and regression trees}.\hskip 1em plus 0.5em
  minus 0.4em\relax Routledge, 2017.

\bibitem{Tolsa2000}
X.~Tolsa, ``Principal values for the cauchy integral and rectifiability,''
  \emph{Proceedings of the American Mathematical Society}, vol. 128, no.~7, pp.
  2111--2119, 2000.

\bibitem{Olson2017PMLB}
R.~S. Olson, W.~La~Cava, P.~Orzechowski, R.~J. Urbanowicz, and J.~H. Moore,
  ``Pmlb: a large benchmark suite for machine learning evaluation and
  comparison,'' \emph{BioData Mining}, vol.~10, no.~1, p.~36, Dec 2017.

\bibitem{romano2021pmlb}
J.~D. Romano, T.~T. Le, W.~La~Cava, J.~T. Gregg, D.~J. Goldberg,
  P.~Chakraborty, N.~L. Ray, D.~Himmelstein, W.~Fu, and J.~H. Moore, ``Pmlb
  v1.0: an open source dataset collection for benchmarking machine learning
  methods,'' \emph{arXiv preprint arXiv:2012.00058v2}, 2021.

\bibitem{Urbanowicz2012}
R.~J. Urbanowicz, J.~Kiralis, N.~A. Sinnott-Armstrong, T.~Heberling, J.~M.
  Fisher, and J.~H. Moore, ``Gametes: a fast, direct algorithm for generating
  pure, strict, epistatic models with random architectures,'' \emph{BioData
  Mining}, vol.~5, no.~1, p.~16, Oct 2012.

\bibitem{ScOt15}
F.~R. Schumacher \emph{et~al.}, ``Genome-wide association study of colorectal
  cancer identifies six new susceptibility loci,'' \emph{Nature
  Communications}, vol.~6, p. 7138, Jul 2015.

\bibitem{das2016next}
S.~Das, L.~Forer, S.~Sch{\"o}nherr, C.~Sidore, A.~E. Locke, A.~Kwong, S.~I.
  Vrieze, E.~Y. Chew, S.~Levy, M.~McGue \emph{et~al.}, ``Next-generation
  genotype imputation service and methods,'' \emph{Nature Genetics}, vol.~48,
  no.~10, pp. 1284--1287, 2016.

\bibitem{loh2016reference}
P.-R. Loh, P.~Danecek, P.~F. Palamara, C.~Fuchsberger, Y.~A~Reshef,
  H.~K~Finucane, S.~Schoenherr, L.~Forer, S.~McCarthy, G.~R. Abecasis
  \emph{et~al.}, ``Reference-based phasing using the haplotype reference
  consortium panel,'' \emph{Nature genetics}, vol.~48, no.~11, pp. 1443--1448,
  2016.

\bibitem{Feltes2019}
B.~C. Feltes, E.~B. Chandelier, B.~I. Grisci, and M.~Dorn, ``Cumida: An
  extensively curated microarray database for benchmarking and testing of
  machine learning approaches in cancer research,'' \emph{Journal of
  Computational Biology}, vol.~26, no.~4, pp. 376--386, 2019, pMID: 30789283.

\bibitem{komorniczak2023problexity}
J.~Komorniczak and P.~Ksieniewicz, ``problexity—an open-source python library
  for supervised learning problem complexity assessment,''
  \emph{Neurocomputing}, vol. 521, pp. 126--136, 2023.

\bibitem{Elizondo2012}
D.~A. Elizondo, R.~Birkenhead, M.~Gamez, N.~Garcia, and E.~Alfaro, ``Linear
  separability and classification complexity,'' \emph{Expert Systems with
  Applications}, vol.~39, no.~9, pp. 7796--7807, 2012.

\bibitem{Eichler2010}
E.~E. Eichler, J.~Flint, G.~Gibson, A.~Kong, S.~M. Leal, J.~H. Moore, and J.~H.
  Nadeau, ``Missing heritability and strategies for finding the underlying
  causes of complex disease,'' \emph{Nature Reviews Genetics}, vol.~11, no.~6,
  pp. 446--450, 2010, iSSN 1471-0064.

\end{thebibliography}



\begin{IEEEbiographynophoto}{Zhendong Sha}
Zhendong Sha 
received the B.Sc. degree from Nanjing University Jinling College, Nanjing, China, in 2013 and the Master degree in Computer Science from the University of New Brunswick, Fredericton, Canada, in 2018. 
Under the supervision of Drs. Ting Hu and Yuanzhu Chen at Queen’s University, his ongoing Ph.D. research covers developing machine learning methods for high-dimensional genetic data with an emphasis on gene-gene interaction and genetic heterogeneity.
Zhendong's primary research area is feature selection using information theory, evolutionary computation and network science. Through his long-term research goals, Zhendong aims to develop a comprehensive analytical framework that will provide clinicians with disease risk predictive models and personalized treatment plans based on a patient's genetic background.
\end{IEEEbiographynophoto}

\begin{IEEEbiographynophoto}{Li Zhu}
Li Zhu received the Bachelor's degree in computing from the Queen's University, Canada, in 2023. He is currently working toward the Master's degree with the University College London, England.
\end{IEEEbiographynophoto}

\begin{IEEEbiographynophoto}{Zijun Jiang}
Zijun Jiang received the Bachelor's degree in computing from Queen's University, Canada, in 2023. He is currently working toward the Master's degree with the University of New South Wales, Australia.
\end{IEEEbiographynophoto}

\begin{IEEEbiographynophoto}{Yuanzhu Chen}
Yuanzhu Chen received the B.Sc. degree from Peking University, China, in 1999, and the Ph.D. degree from Simon Fraser University, Canada, in 2004. He has been a Professor of computing science since 2005 and is currently affiliated with School of Computing, Queen’s University.  From 2004 to 2005, he was a Postdoctoral Researcher with Simon Fraser University. In 2005, Dr. Chen joined Memorial University as a tenure-track Assistant Professor. While at Memorial, he was the Deputy Head for Undergraduate Studies from 2012 to 2015, the Deputy Head for Graduate Studies from 2016 to 2019, and Department Head from 2019 to 2021. He then joined Queen’s School of Computing in 2021.  Dr. Chen's research interests include complex networks, computer networking, online social networks, mobile computing, graph theory, Web information retrieval, and evolutionary computation, with funding from national agencies and various university programs and awards. He was a recipient of the President’s Award for Distinguished Teaching in 2018. 
\end{IEEEbiographynophoto}

\begin{IEEEbiographynophoto}{Ting Hu}
Ting Hu is currently an Associate Professor from the School of Computing at Queen's University in Canada.
She received a B.Sc. degree in applied mathematics and an M.Sc. degree in computer science from Wuhan University, China, in 2003 and 2005, respectively, and a Ph.D. degree in computer science from Memorial University, Canada, in 2010.
She received her postdoctoral training in computational genetics at the Geisel School of Medicine, Dartmouth College in USA. She was previously an Assistant Professor at the Department of Computer Science, Memorial University in Canada. Her research interests include computational intelligence, machine learning, and bioinformatics.
She has served on the editorial board of international journals including Genetic Programming and Evolvable Machines, Entropy, Genes, and has chaired international conferences including European Conference on Genetic Programming, Genetic and Evolutionary Computation Conference GP track, European Conference on Evolutionary Computation, Machine Learning and Data Mining in Bioinformatics.
\end{IEEEbiographynophoto}

\vfill

\end{document}